%% file: main.tex
\documentclass[floats,floatfix,showpacs,amssymb,prd,twocolumn,superscriptaddress,nofootinbib,nolongbibliography,reprint]{revtex4-2}

\usepackage{amssymb,amsmath,verbatim,mathtools,needspace,enumitem,etoolbox,graphicx,physics,microtype,afterpage,bigints}
\usepackage{gensymb}

\usepackage[dvipsnames, usenames, table]{xcolor}
\usepackage{multirow}
\usepackage{rotating}
\definecolor{linkcolor}{rgb}{0.0,0.3,0.5}
\usepackage[unicode, colorlinks=true, linkcolor=linkcolor, citecolor=linkcolor, filecolor=linkcolor,urlcolor=linkcolor, pdfusetitle]{hyperref}
\usepackage[all]{hypcap}
\usepackage[T1]{fontenc}
\usepackage[utf8]{inputenc}
\usepackage{hhline}

\usepackage[normalem]{ulem}
\usepackage{booktabs} 
\usepackage{longtable}

\renewcommand{\arraystretch}{1.4}
\newcommand{\ssim}{\mathchar"5218\relax\,}

\newcommand{\chieff}{\chi_{\mathrm{eff}}}
\newcommand{\chip}{\chi_{\mathrm{p}}}

\renewcommand{\vec}[1]{\mathbf{#1}}

\def\apj{\rm{Astrophys.~J.}}
\def\apjl{\rm{Astrophys.~J.~ Lett.}}
\def\apjs{\rm{Astrophys.~J.~Suppl.~S.}}

\def\mnras{\rm{Mon.~Not.~R.~Astron.~Soc.}}
\def\prd{\rm{Phys.~Rev.~D}}
\def\prl{\rm{Phys.~Rev.~Lett.}}
\def\prx{\rm{Phys.~Rev.~X}}
\def\prr{\rm{Phys.~Rev.~Res.}}
\def\pasa{\rm{PASA}}

\def\cqg{\rm{Class.~Quantum~Gravity}}
\def\lrr{\rm{Living~Rev.~Relativ.}}

\newcommand{\bham}{\affiliation{School of Physics and Astronomy \& Institute for Gravitational Wave Astronomy, \\University of Birmingham, Birmingham, B15 2TT, UK}}
\newcommand{\dallas}{\affiliation{Department of Physics, The University of Texas at Dallas, Richardson, Texas 75080, USA}}
\newcommand{\milan}{
\affiliation{Dipartimento di Fisica ``G. Occhialini'', Universit\'a degli Studi di Milano-Bicocca, Piazza della Scienza 3, 20126 Milano, Italy}
\affiliation{INFN, Sezione di Milano-Bicocca, Piazza della Scienza 3, 20126 Milano, Italy}
}

\newcommand\orcidlink[1]{\href{https://orcid.org/#1}{$\!$\includegraphics[scale=0.006]{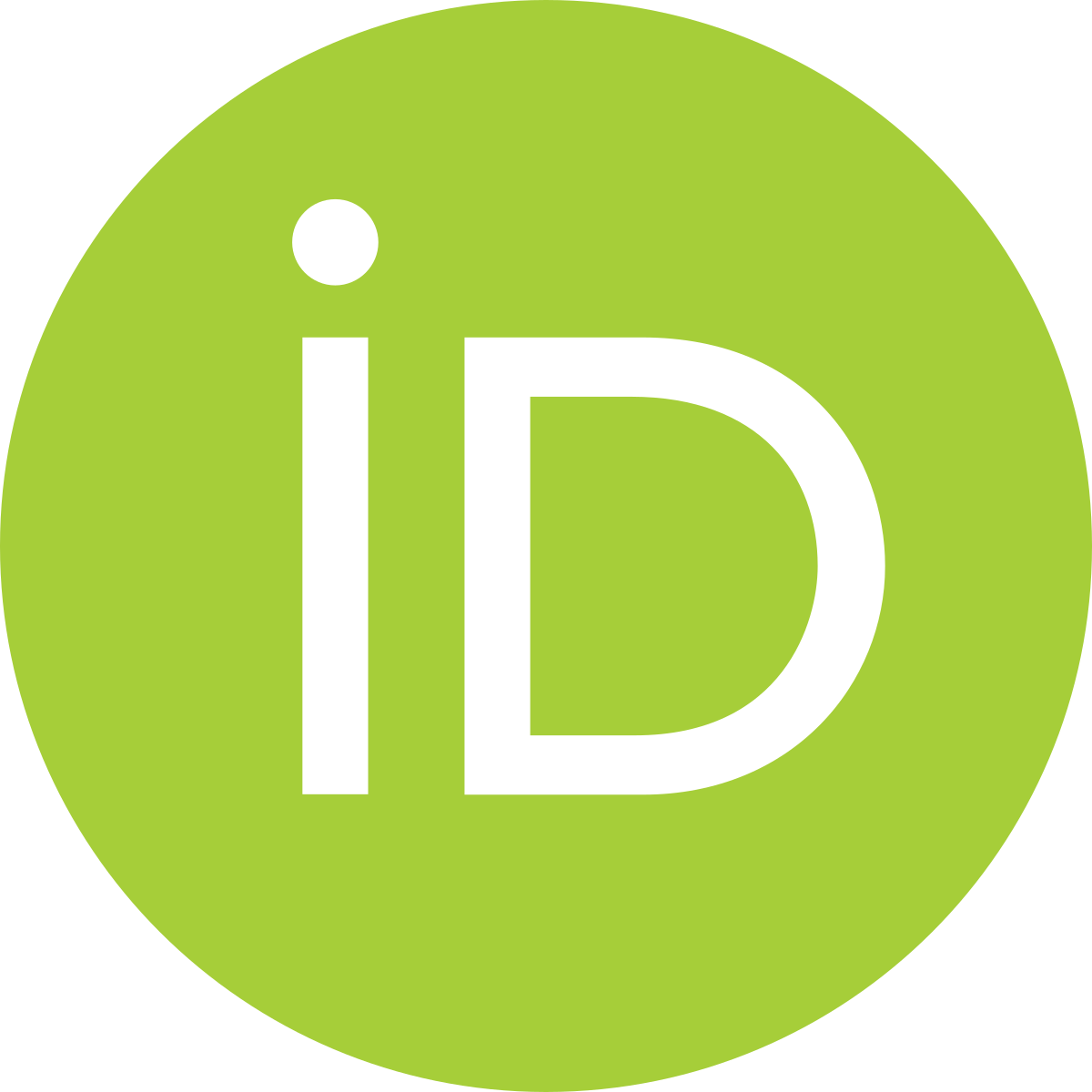} $\!\!$}}

\begin{document}

\title{Constraining black-hole binary spin precession and nutation  \\ 
with sequential prior conditioning}

\author{Daria Gangardt \orcidlink{0000-0001-7747-689X}}

\email{ddg672@star.sr.bham.ac.uk}
\bham

\author{Davide Gerosa \orcidlink{0000-0002-0933-3579}}
\milan \bham

\author{Michael Kesden \orcidlink{0000-0002-5987-1471}}
\dallas

\author{Viola De Renzis \orcidlink{0000-0001-7038-735X}}
\milan

\author{Nathan Steinle \orcidlink{0000-0003-0658-402X}}
\bham

 \pacs{}

\date{\today}

\begin{abstract}

We investigate the detectability of sub-dominant spin effects in merging black-hole binaries using current gravitational-wave data. Using a phenomenological model that separates the spin dynamics into precession (azimuthal motion) and nutation (polar motion), we present constraints on the resulting amplitudes and frequencies. We also explore current constraints on the spin morphologies,  indicating if binaries are trapped near spin-orbit resonances. We dissect such weak effects from the signals using a sequential prior conditioning approach, where parameters are progressively re-sampled from their posterior distribution. This allows us to investigate  whether the data contain additional information beyond what is already provided by quantities that are better measured, namely the masses and the effective spin. For the current catalog of events, we find no significant measurements of weak spin effects such as nutation and spin-orbit locking. We synthesize a source with a high nutational amplitude and show that near-future detections will allow us to place powerful constraints, hinting that we may be at the cusp of detecting spin nutations in gravitational-wave data.

\end{abstract}

\maketitle

\section{Introduction}
\label{sec:Intro}

The first three observing runs of the Advanced LIGO-Virgo-KAGRA network have delivered %
 $\ssim 90$ gravitational wave (GW) events~\cite{2019PhRvX...9c1040A, 2021arXiv210801045T, 2021PhRvX..11b1053A, 2021arXiv211103606T}; additional significant triggers from the same stretches of data have also been reported~\cite{2019PhRvD.100b3007Z, 2020PhRvD.101h3030V, 2020ApJ...891..123N, 2021PhRvD.104f3030Z, 2021ApJ...922...76N, 2021arXiv211206878N}. Of these observations, about $80$ of them are mergers of binary black holes (BBHs).

GWs give us an insight into the intrinsic properties of the merging BBHs. If the component black holes are spinning, their spins will interact with both the binary's orbital angular momentum and each other, causing the system to  precess~\cite{1994PhRvD..49.6274A}. 
The short duration of GW signals and the subtle effects of spin precession on the gravitational waveform mean that, so far, a confident detection of spin precession in a single BBH event remains elusive~\cite{2021arXiv211103634T}. GW200129\_065458 is a possible exception, with Refs.~\cite{2021arXiv211211300H, 2022PhRvL.128s1102V} claiming that a re-analysis of the data with different waveform models
can indeed unveil signatures of orbital-plane precession. %

Stronger evidence was reported at the population level, with the current catalog indicating that some spin precession is necessary to explain the data at $>99$\% level of confidence~\cite{2021arXiv211103634T}.  Constraining spin precession in GW events informs us on how the BBHs formed, with  different formation channels predicting different BBH spin properties~\cite{2010CQGra..27k4007M, 2013PhRvD..87j4028G, 2016ApJ...832L...2R, 2017PhRvD..96b3012T, 2017ApJ...846...82Z, 2017MNRAS.471.2801S, 2018PhRvD..98h4036G, 2022MNRAS.511.5797M, 2021PhRvD.103h3021W, 2021PhRvD.103f3032S}. 
Furthermore, several rare but interesting spin configurations are predicted to leave a unique signature on the GW strain, including the spin-orbit resonances~\cite{2004PhRvD..70l4020S,2014CQGra..31j5017G,2015PhRvD..92f4016G,2016MNRAS.457L..49C,2018PhRvD..98h3014A, 2022PhRvL.128c1101V}, transitionally precessing binaries~\cite{1994PhRvD..49.6274A,2017PhRvD..96b4007Z}, instabilities~\cite{2015PhRvL.115n1102G,2016PhRvD..93l4074L,2020PhRvD.101l4037M,2021PhRvD.103f4003V}, and widely nutating sources (also referred to as ``spin flips'')~\cite{2015PhRvL.114n1101L,2016PhRvD..93d4031L,2019CQGra..36j5003G}.
As GW detectors become more and more sensitive and the population of BBHs grows, the measurability of spin effects is bound to improve for both single-event outliers and the entire population of detected sources.

The high dimensionality of the BBH spin parameter space and the weak effect of spin precession on the waveform have prompted state-of-the-art analyses to report results using a limited number of combined estimators that are believed to encode the majority  of the available information. Most often these are the so-called effective spin parameter $\chieff$~\cite{2001PhRvD..64l4013D, 2008PhRvD..78d4021R} and the  precession parameter $\chip$~\cite{2015PhRvD..91b4043S, 2021PhRvD.103f4067G}. 

Let us consider two black holes of masses $m_1$ 
 and $m_2$, %
  mass ratio $q \equiv m_2/m_1 \leq 1$, total mass $M=m_1+m_2$, spin magnitudes $S_i = m_i^2\chi_i$ (hereafter we set $c=G=1$), and  dimensionless Kerr parameters  $\chi_i\in[0,1]$. 
The effective spin parameter $\chieff$ is defined as
\begin{equation*}
    \chieff \equiv \left [ (1+q)\vec{S}_1 + \left(\frac{1+q}{q} \right)\vec{S}_2\right]\cdot \frac{\hat{\vec{L}}}{M^2},
\end{equation*}
where $\vec{L}$ is the Newtonian orbital angular momentum. The quantity $\chieff$ is a constant of motion at 2 Post-Newtonian (PN) order  in spin precession~\cite{2008PhRvD..78d4021R}, making it useful to describe the full evolution of BBHs from large separations to merger.
The effective precession parameter $\chip$ is proportional to the variation of the direction of the angular momentum $|d \hat{\vec{L}} /dt|$~\cite{2015PhRvD..91b4043S} and can be averaged over the binary's spin precession cycle to obtain a more resilient estimator~\cite{2021PhRvD.103f4067G}. Several other spin-precession parametrizations have been proposed, notably including the precession signal to noise ratio $\rho_P$~\cite{2020PhRvD.102d1302F} and a two-dimensional precession vector $\boldsymbol{\chi}_\perp$~\cite{2021PhRvD.103h3022T}.
 While well motivated, we argue these parameters make the interpretation of the systems  opaque, somewhat obscuring the underlying dynamical picture.

Taking a step back from the current effective-spin ideas, some of the authors previously proposed alternative estimators that stems directly from PN dynamics~\cite{2015PhRvL.114h1103K,2015PhRvD..92f4016G,2021PhRvD.103l4026G}. In Ref.~\cite{2021PhRvD.103l4026G} we split the motion of the orbital angular momentum around the total angular momentum into its nutational (polar) and precessional (azimuthal) components,
and use the resulting frequencies and amplitude as indicators of BBH spin precession. 
In Refs.~\cite{2015PhRvL.114h1103K,2015PhRvD..92f4016G} we illustrated how BBHs can be divided into mutually exclusive ``morphologies'' based on the shape of their precession cones. %
These spin morphologies reduce to the known spin-orbit resonances~\cite{2004PhRvD..70l4020S} in their zero-amplitude limit, thus generalizing the more stringent co-planarity condition of the three spin vectors $\vec{S}_{1,2}$, and $\vec{L}$. %
Both the phenomenological amplitudes and frequency parameters~\cite{2021PhRvD.103l4026G} as well as the spin morphologies~\cite{2015PhRvL.114h1103K,2015PhRvD..92f4016G} have yet to be constrained using the data from current GW event catalogs.

Much like the effective spins, our spin-precession estimators also depend on the masses and spin components of the BBHs in non-trivial ways. The resulting Bayesian posteriors are highly correlated, which can make disentangling effects and interpretation of data somewhat  challenging. This is especially true for weak observables such as those due to spin precession, where the data are only mildly informative. A pertinent question to ask in this context is therefore the following: 
\begin{quote} 
Are constraints on the precession parameters providing information beyond what is already encoded in the  other observables?
\end{quote}

We tackle this point using sequential prior conditioning.
In brief, constructing a conditional prior implies combining the posterior samples of the parameter(s) we are interested in with the uninformative prior distributions of the other parameters. An example of such a procedure in GW astronomy can be seen in Fig. 10 of Ref.~\cite{2021arXiv211103606T}, where the $\chip$ priors have been conditioned on $\chieff$. Prior conditioning is an effective strategy to highlight parameter correlations and show to what extent a given estimator uncovers new information from the data. 
A more common approach to identifying features in the data is that of calculating odds ratios between analyses where the putative features are included/excluded. While this readily allows one to constrain the joint effect of spin precession and nutation (one needs to compare inference runs with precessing spins against control cases where spins are assumed to be aligned, e.g. \cite{2020PhRvL.125j1102A, 2020PhRvR...2d3096P, 2021arXiv211110455H, 2021PhRvD.103b4029C}), current signal models do not isolate one from the other. Our approach aims to be complimentary and seeks to investigate if using more phenomenological parameters can uncover additional information.%

Among the intrinsic parameters of a GW event, we expect the BBH masses and the effective spin parameter $\chieff$ to have a large influence on our posteriors, with the spin precession estimators providing a subdominant contribution. Therefore, events with precession parameters constrained away from their priors conditioned on both the masses and $\chieff$ would provide smoking-gun evidence that new information about the event is being revealed. 

In this paper, we systematically employ sequential prior conditioning to investigate if and how the dynamics-based estimators of Refs.~\cite{2015PhRvL.114h1103K,2015PhRvD..92f4016G,2021PhRvD.103l4026G} can be used to constrain BBH spin precession measured in current GW data. %
 In Sec.~\ref{estimators} we briefly review the formulation of the precession/nutation amplitudes and frequencies, as well as  the spin morphologies. Section~\ref{conditional_priors} details the methodology required to sequentially condition priors on measured parameter posteriors. %
In Sec.~\ref{sec:plots} we present our results using data from the current GW catalog. Perhaps unsurprisingly, current evidence is weak. In Sec.~\ref{injection} we present a preliminary analysis from synthetic LIGO/Virgo data and highlight prospects for future observations. Finally, in Sec.~\ref{sec:conclusions} we draw our conclusions. Some more detailed results are postponed to Appendices \ref{injapp} and \ref{app}.  %

\section{Spin precession estimators}
\label{estimators}

\subsection{Five parameters from the decomposition of precession and nutation}

Our estimators rely on the PN precession-averaged approach first developed in Refs.~\cite{2015PhRvL.114h1103K,2015PhRvD..92f4016G}  and  explored at 
length by both ourselves and other authors~\cite{2017CQGra..34f4004G,2017PhRvD..96b4007Z,2017PhRvD..95j4004C,2019CQGra..36j5003G,2019PhRvD.100b4059K, 2020CQGra..37v5005R,2021PhRvD.103l4026G,2021arXiv210711902J,2021arXiv210610291K}. In particular, the spin dynamics 
 is decomposed into the azimuthal (``precession'') and polar (``nutation'') %
motions of the Newtonian orbital angular momentum $\vec{L}$.%
 We only tackle the secular evolution of the spins, which rely on  orbit-averaged  equations of motions \cite{2008PhRvD..78d4021R}. This implies that we are not sensitive to the dynamics happening on the short orbital timescale (which itself includes nutations, see e.g. \cite{2015PhRvD..92j4028O}).

The vector $\vec{L}$ moves around $\vec{J}$ with an azimuthal frequency $\Omega_L$ and an %
opening angle $\theta_{L}$. The expressions for these quantities can be computed analytically at 2PN and are reported in Eqs.~(29)  and (6)  of Ref.~\cite{2015PhRvD..92f4016G}, respectively. In the limit where there is no nutation, $\Omega_L$ and $\theta_{L}$ are constant on the precession timescale and only slowly evolve on the radiation-reaction timescale. 
This simple limit, which we refer to as ``regular precession''~\cite{2021PhRvD.103l4026G}  following the classical nomenclature used for dynamical systems~\cite{1969mech.book.....L}, corresponds to $\vec{L}$ evolving around $\vec{J}$  on a cone with a fixed opening angle and a constant angular velocity.

Regular precession requires fine-tuned conditions~\cite{2021PhRvD.103l4026G}. %
In the generic case, the frequency $\Omega_L$ and the opening angle $\theta_{L}$ are not constant, but evolve on a timescale that is comparable to that taken by $\vec{L}$ to evolve around $\vec{J}$, causing a nutational motion.
In the PN regime where radiation-reaction can be safely assumed to be a slow, quasi-adiabatic process, Refs.~\cite{2015PhRvL.114h1103K,2015PhRvD..92f4016G} showed that the entire dynamics can be described using only one parameter (the underlying expressions are akin to the familiar effective-potential method used in Keplerian dynamics). This parameter can be chosen to be magnitude of the total spin %
$S = |\vec{S}_1 + \vec{S}_2|$~\cite{2015PhRvL.114h1103K,2015PhRvD..92f4016G}, (but see Refs.~\cite{2017CQGra..34f4004G,2021arXiv210610291K} for other suitable parametrizations). The motion of $S$ is periodic and ranges %
between a maximum $S_{+}$ and minimum $S_{-}$ with velocity $dS/dt$; %
this can also be computed analytically at 2PN, see Eq.~(26) in Ref.~\cite{2015PhRvD..92f4016G}. A full $S$ oscillation is completed in a period $\tau=2 \int_{S_-}^{S_+} |dt/dS| dS$ and takes place on a frequency $\omega = 2\pi/\tau$.

An explicit parametrization of the spin evolution allows for a consistent averaging of physical properties over a nutation cycle. For a quantity of interest $X$, its precession average is simply given by
$\langle X\rangle = (2/\tau) \int_{S_-}^{S_+}  X(S)|dt/dS| dS$\,.
In the generic case where both precession and nutation are present, the motion of $\bf {L}$ can thus be described by the \emph{averaged} values of the opening angle $\langle \theta_L \rangle$ and the precession frequency $\langle \Omega_L \rangle$.  The variations of these quantities over a nutation cycle can be captured by the differences $\Delta\theta_L = [\theta_{L}(S_+) - \theta_{L}(S_-)]/2$ and $\Delta\Omega_L = [\Omega_{L}(S_+) - \Omega_{L}(S_-)]/2$.

Summarizing these efforts, %
Gangardt~\&~Steinle $et~al.$~\cite{2021PhRvD.103l4026G} %
proposed to dissect the BH binary dynamics using five phenomenological parameters that, together, describe the joint precessional and nutational motions. These are: %

\begin{enumerate}[label=(\roman*)]
    \item The precession amplitude $\langle \theta_L \rangle$.
    \item The precession frequency $\langle\Omega_L\rangle $.
    \item The nutation amplitude $\Delta\theta_L$.
    \item The nutation frequency $\omega$.
    \item The variation of the precession frequency $\Delta\Omega_L$.
    \end{enumerate}
The astrophysical consequences of these five parameters are explored in Ref.~\cite{2021MNRAS.501.2531S} and Ref.~\cite{2022arXiv220600391S} for supermassive and stellar-mass BHs, respectively.

\subsection{Spin morphologies}
    
A complementary categorization that stems directly from the precession-averaged formalism is that of the spin morphologies. These generalize the notion of the spin-orbit resonances~\cite{2004PhRvD..70l4020S}, which are  non-trivial configurations where nutation vanishes and the four vectors $\bf{S}_1$, $\bf{S}_2$, $\bf{L}$, and $\bf{J}$ remain coplanar (see Refs.~\cite{2013PhRvD..87j4028G,2014PhRvD..89l4025G,2014CQGra..31j5017G,2016PhRvD..93d4071T,2016MNRAS.457L..49C,2018PhRvD..98h3014A,2022PhRvL.128c1101V} for some of the numerous explorations on the topic).  There are two families of resonant solutions, characterized by the only two possible configurations that define co-planarity: $\Delta\Phi=0$ and $\Delta\Phi=\pi$, where $\Delta \Phi$ is the angle between the projections of the two spins onto the orbital plane. Starting from these configurations of regular precession, the entire parameter space of spinning BH binaries can be divided into three mutually exclusive classes where:
\begin{enumerate}[label=(\roman*)]
    \item Binaries librate in the vicinity of the $\Delta\Phi=0$ resonance (L$0$).   
        \item Binaries circulate freely far from either of the two resonances (C).
    \item Binaries  librate in the vicinity of the $\Delta\Phi=\pi$ resonance (L$\pi$).   
\end{enumerate}
Crucially, not all morphologies are available to all binaries: the parameters that are constant on the spin-precession timescale ($q$, $J$, $S_1$, $S_2$, $r$, $\chi_{\rm eff}$) can restrict sources to only having certain morphologies~\cite{2015PhRvD..92f4016G}. The secular evolution of $J$ and $r$ on the radiation-reaction timescale can cause transitions between the different classes. The spin morphology is thus a quantity that classifies the spin dynamics while being constant on the spin-precession timescale. In the LIGO context, this feature could potentially be exploited to probe BH binary formation channels~\cite{2018PhRvD..98h4036G,2019PhRvD..99j3004G,2022arXiv220600391S}. %

\section{Dissecting information}
\label{conditional_priors}

\subsection{Conditional priors}

GW parameter estimation is typically performed within the framework of Bayesian statistics, which explicitly require assuming a prior distribution on the targeted parameters. The standard analyses~\cite{2019PhRvX...9c1040A, 2021PhRvX..11b1053A, 2021arXiv211103606T} assume a prior that is uniform in $m_1$ and $m_2$ (though with cuts in this 2-dimensional parameter space that are informed by the output of the preceding search pipelines), uniform in the spin magnitudes $\chi_1$ and $\chi_2$, and isotropic in the spin directions.
This is often referred to as the ``uninformative'' prior.\footnote{While we use this term for consistency with the literature on the topic,  it is a misnomer because the choices behind these prior assumptions are subjectively elicited.}

Starting from these prior assumptions, stochastic sampling is used to obtain the posterior distribution of the binary parameters. %
The posterior conveys our best knowledge of the observed BHs.
Using public samples from Refs.~\cite{2020MNRAS.499.3295R,2021arXiv210801045T, 2021PhRvX..11b1053A, 2021arXiv211103606T},  
we select  the BBH events that have a probability of astrophysical origin $>0.5$ (unless the events are listed in the GWTC-1 catalog, in which case we use all of them). We include BBH events with secondary masses above $2.2M_{\odot}$ in the source frame; the chosen neutron star threshold reflecting the mass distribution obtained from pulsar observations~\cite{2018MNRAS.478.1377A}. Where possible, we use the default samples that combine equally parameter estimation results from the Phenom and EOB waveform families (c.f.~\cite{2020MNRAS.499.3295R,2021arXiv210801045T, 2021PhRvX..11b1053A, 2021arXiv211103606T} and  references therein). For events where such combined results are unavailable, we use samples from the Phenom waveform family only. We use priors that are uniform in comoving volume and source-frame time.

We re-cast prior and posterior distributions for each event in terms of the five parameters of Ref.~\cite{2021PhRvD.103l4026G} and the spin morphologies of Ref.~\cite{2015PhRvD..92f4016G} using the \textsc{precession} code~\cite{2016PhRvD..93l4066G}. The necessary quantities for this procedure are the masses, the spins (both magnitudes and directions), and the PN separation of the binary $r$ at the reference frequency of the parameter estimation. LIGO/Virgo parameter estimation samples are reported at a 
 reference frequency of 20 Hz for all events except GW190521, which has a reference frequency of 11 Hz. For each sample, we estimate the orbital separation $r$ using Eq.~(4.13) of Ref.~\cite{1995PhRvD..52..821K}; this conversion needs to be performed using detector-frame masses.

\begin{figure}
    \includegraphics[width=\columnwidth]{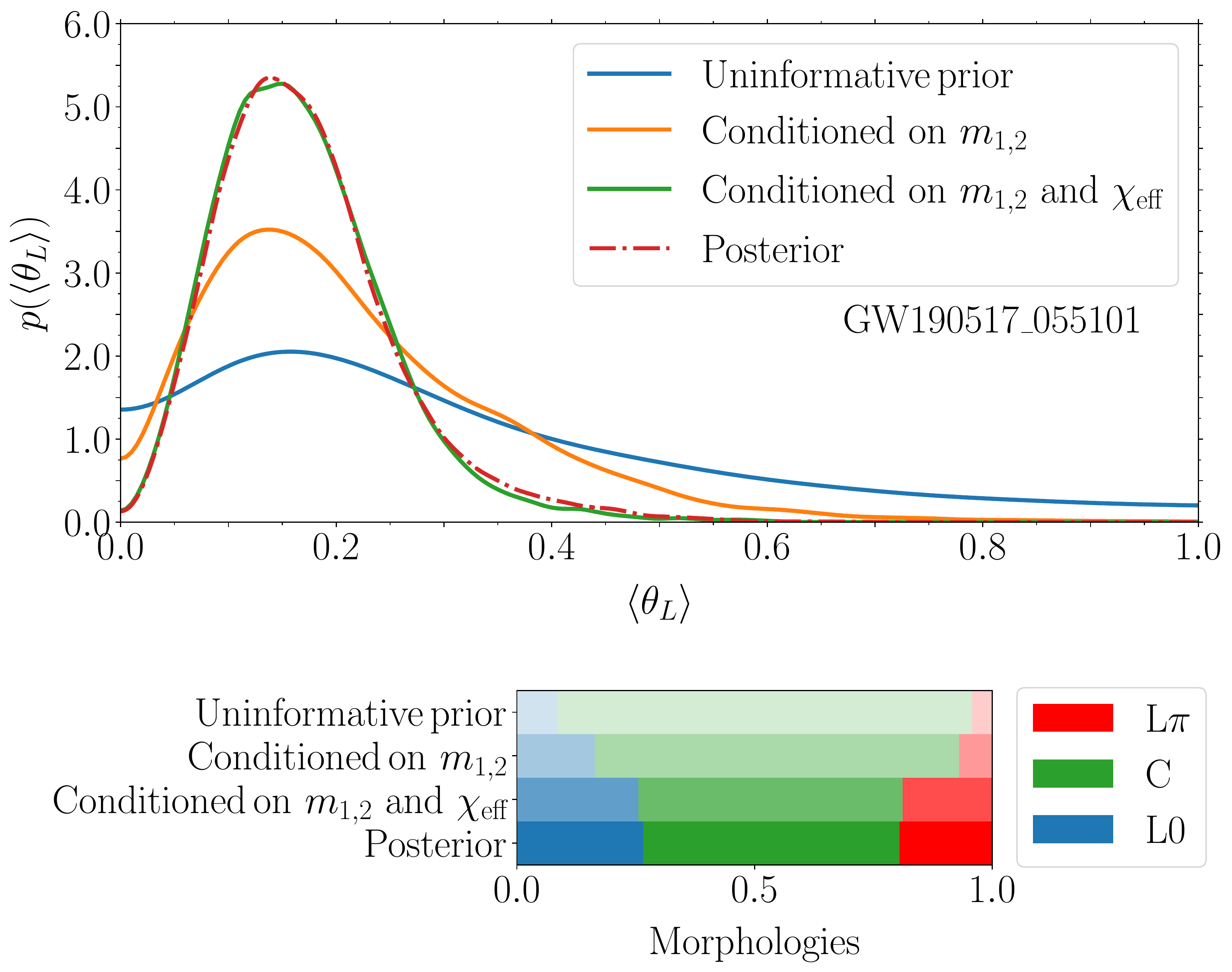}
    \caption{Prior and posterior distributions for the precession amplitude $\langle \theta_L \rangle$ and the spin morphology of GW190517\_055101. In both cases we show the uninformative prior, the prior conditioned on the masses, the prior conditioned both masses and effective spin, and the posterior. For the case of the continuous parameter $\langle \theta_L \rangle$, distributions are illustrated using kernel density estimation (top panel). For the case of the spin morphology (bottom panel), we show the fraction of samples in each of the three mutually excluding classes L$\pi$, C, and L$0$. In both cases, the prior conditioned on both masses and $\chieff$ is nearly identical to the posterior distribution, indicating that  measurements of those parameters already constrain the precession estimators almost entirely.}
    \label{fig:single_event}
\end{figure}

\begin{figure}[ht]
    \centering
    \includegraphics[width =\linewidth]{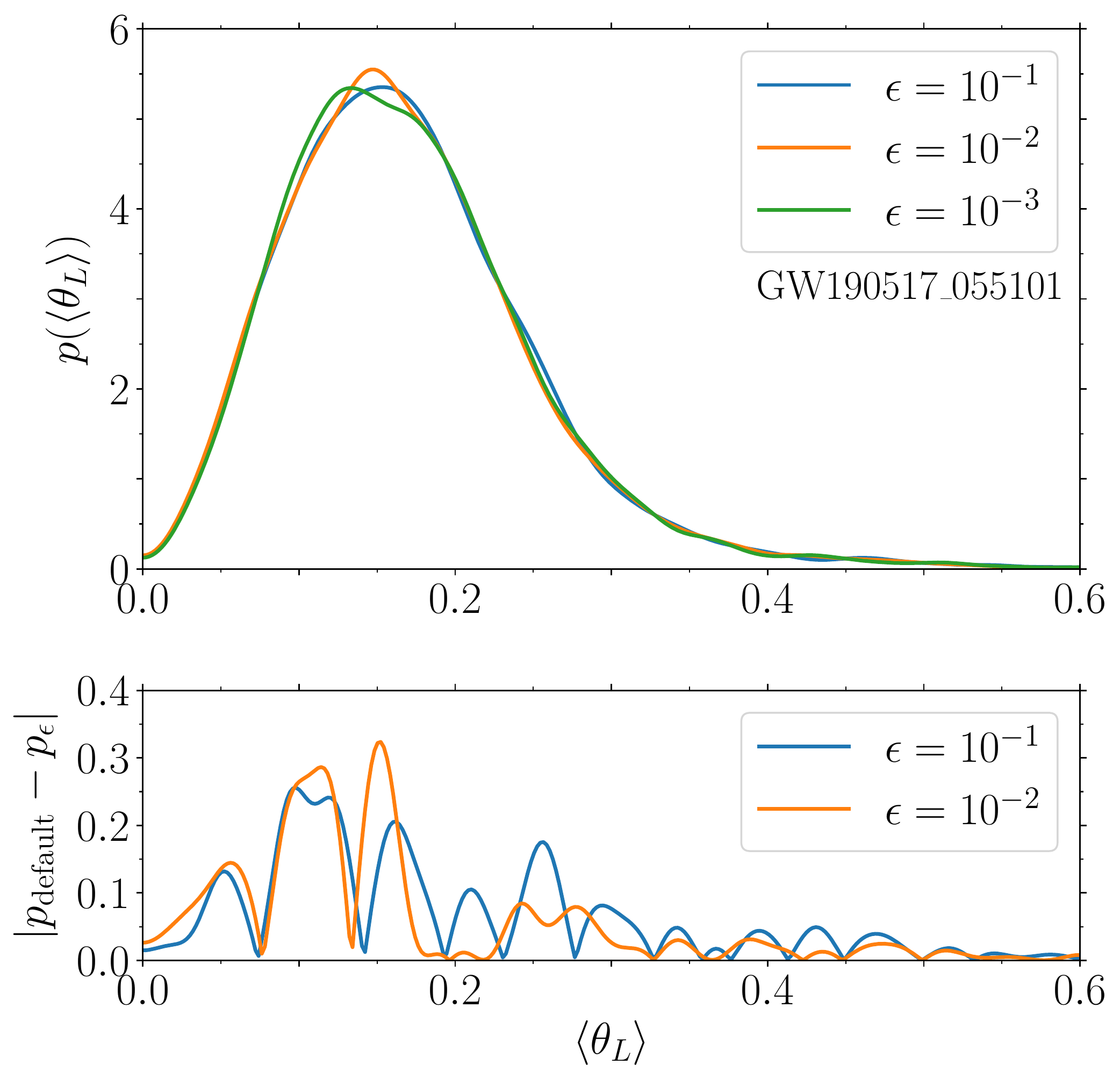}
    \caption{Numerical threshold when conditioning on masses and effective spin. The top panel shows the resulting precessional amplitudes $\langle \theta_L \rangle$ of GW190517\_055101 for three different thresholds $\epsilon= 10^{-1}$ (blue), $10^{-2}$ (orange) and $10^{-3}$ (green). The bottom panel shows  residuals against the conditioned prior obtained with our default threshold ($\epsilon = 10^{-3}$). }
    
    \label{fig:fig2}
\end{figure}

When dealing with weak effects in Bayesian statistics, one needs to worry about whether the observed  features are data- or prior-driven (see Refs.~\cite{2017PhRvL.119y1103V,2022PASA...39...25R} for GW explorations on this point). Inference on degenerate parameters prompts the question on whether there is truly new information that can be extracted, or whether the data are  already saying everything there is to say. In our case, spin precession has a subdominant effect on the waveform and some regions of the parameter space of the spin degrees of freedom are only available to binaries with certain values of the parameters. We address this issue with sequential prior conditioning, which increases the granularity between prior and posterior, hopefully highlighting where the targeted effects come into play. Prior conditioning has been used in previous analyses when comparing the effective precession parameter $\chip$ posteriors to priors conditioned on $\chieff$~\cite{2019PhRvX...9c1040A, 2021PhRvX..11b1053A, 2021arXiv210801045T, 2021arXiv211103606T}. %

The masses are generically easier to constrain than the spins. Therefore, we first condition our spin inference on the measured values of $m_1$ and $m_2$ (or, equivalently, total mass and mass ratio). This is straightforward to implement because the uninformative prior  assumes that masses and spins are uncorrelated~\cite{2019PhRvX...9c1040A, 2021PhRvX..11b1053A, 2021arXiv210801045T, 2021arXiv211103606T}. One can %
simply take  the marginalized posterior distributions of the two masses and combine them with random samples drawn from the uninformative prior for the spins.

Next, it is well known that among the spin degrees of freedom, the combination $\chieff$~\cite{2001PhRvD..64l4013D,2008PhRvD..78d4021R} is better measured because it affects the length of the waveform. We thus wish to build a prior that is conditioned on all three parameters $m_1$, $m_2$, and $\chieff$. %
The implementation here is less trivial because the uninformative prior is posed on $m_i$ and ${\bf S}_i$ separately, resulting in a prior on   $\chi_{\rm eff}$ that depends on the event-based cuts. We adopt the following numerical approach.  For each mass sample in the posterior distribution, we extract a random draw from the uninformative spin prior  and compute the resulting $\chieff$. We then compare this against the posterior's $\chieff$ and accept the draw if their absolute difference is below a specified threshold $\epsilon=10^{-3}$. The process is iterated, individually for each sample, until a matching draw is found. 
 
We thus construct four distributions of our spin-precession estimators: 
\begin{enumerate}[label=(\roman*)]
    \item The uninformative prior.
     \item The prior conditioned on the $m_{1}$ and $m_2$ posteriors.
     \item The prior conditioned on the $m_1$, $m_2$ and $\chieff$ posterior.
     \item The posterior. \end{enumerate}
An example of such sequential  conditioning is reported in Fig.~\ref{fig:single_event} for GW190517\_055101, which is an event with a relatively high value of $\chieff$ ($\chieff = 0.54_{-0.19}^{+0.19}$). We show probability distributions for two of our spin estimators, the precession amplitude $\langle \theta_L \rangle$ and the spin morphology. %
This highlights what information on spin precession remains present in the data as one goes from prior to posterior across the two conditionings - we see for both estimators, the prior distributions become increasingly similar to the posterior distribution.

Figure \ref{fig:fig2} shows a convergence study for the numerical threshold $\epsilon$. We test three different thresholds for the nutational 
amplitude of GW190517\_055101. The resulting $\chieff$ and mass conditioned prior distributions of $\langle \theta_L \rangle$ show differences of $\lesssim 0.2$  between our two higher-resolution runs without evident systematics.  We have also tested the convergence of all the other estimators and report similar accuracy. GW190517\_055101 is the event whose $\chieff$ posterior distribution is relatively well constrained furthest from $\chieff = 0$ (where the uninformative priors tend to be the largest), thus we expect it to be the most sensitive to thresholds in $\epsilon$, making results in Fig.~\ref{fig:fig2} conservative and justifying our chosen default threshold of $\Delta \chieff = 10^{-3}$.

\subsection{Distance between probability distributions} \label{distance}

Some of the more common choices used to compute the difference between two probability distributions include the Kullback-Leibler  divergence, its symmetrized extension by Jensen and Shannon~\cite{lin1991divergence}, and the Hellinger distance~\cite{hellinger1909neue}.
Here we employ the latter
because it satisfies some very desirable properties including symmetry and unit range (cf. Ref.~\cite{2021PhRvD.104h3008M} for a physicists summary).  The Hellinger distance between two continuous probability distributions $p(x)$ and $q(x)$ is defined as
\begin{equation}
    d_H^2 =   1 - \int{\mathrm{d}x \sqrt{p(x) q(x)}}.
\end{equation}
For the discrete case where $p$ and $q$ can take $N$ values (as in the case of the spin morphologies) one instead has
\begin{equation}
    d_H^2 =   1 - \sum_{i=1}^N \sqrt{p_i q_i}.
\end{equation}
The Hellinger distance can take values in the range $[0, 1]$ where $d_H=0$ for two identical distributions and $d_H=1$ whenever the supports of $p$ and $q$ do not overlap. For comparison, the Hellinger distance between two identical normal distributions that are offset by $n$ standard deviations is $d_H^2 = 1- \exp (-n^2/8)$, which implies $d_H\simeq 0.12$ for a 1-$\sigma$ difference.

\section{Inference from current data}
\label{sec:plots}

We now examine the distributions of our estimators across the current GW catalog.  First, we concentrate on a single event for illustrative purposes. 

\begin{figure*}[tb]
    \centering
    \includegraphics[width = 0.8\textwidth]{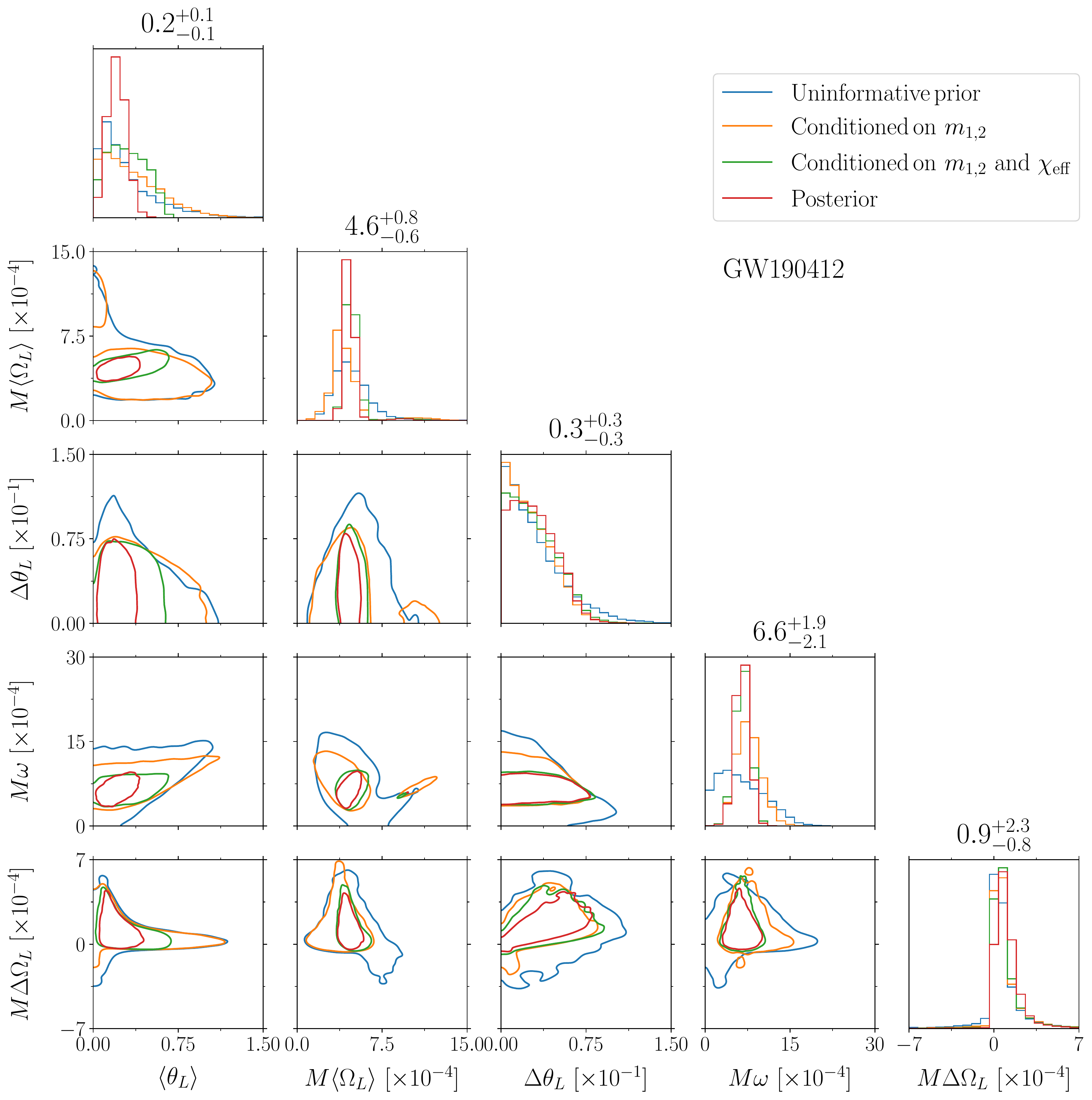}
    \caption{Uninformative prior (blue), conditioned priors (orange, green) and posterior (red) distributions of the five phenomenological parameters describing the joint precessional/nutational dynamics for event GW190412. Joint 2D distributions show $90$\% and $50$\% confidence levels.}
   
    \label{fig:fig3}
\end{figure*}

\subsection{Key behavior of the nutation parameters}

Figure \ref{fig:fig3} shows the distributions of the five precessional and nutational parameters for GW190412~\cite{2020PhRvD.102d3015A}. This is a BH merger with a mass ratio $q = 0.28^{+0.26}_{-0.03}$ that is confidently constrained away from unity and an effective spin $\chieff=0.25^{+0.08}_{-0.11}$ that is confidently constrained away from zero. This makes GW190412 ideal to showcase our sequential conditioning approach. 

The precessional amplitude $\langle \theta_L \rangle$ has a broad prior distribution, partly because the uninformative priors on masses and spins lead to a preference toward small values of $\langle \theta_L \rangle$. Conditioning our priors on the mass parameters results in a broader distribution, retrospectively showing that the uninformative prior's preference for lower $\langle \theta_L \rangle$ was indeed due to the uninformative mass priors.
Lower values of $q$ allow for larger values of $\langle \theta_L \rangle$~\cite{2021PhRvD.103l4026G}, which can be seen in the broadening of the $\langle \theta_L \rangle$ prior once it is conditioned on GW190412's mass parameters.
When we condition our priors on both the masses and $\chieff$, the range of the distribution becomes considerably smaller and is constrained away from $\langle \theta_L \rangle = 0$. GW190412 has a posterior distribution that prefers positive and non-zero values of $\chieff$ and was reported to show mild evidence of spin precession~\cite{2020PhRvD.102d3015A,2020PhRvL.125j1103G}, in agreement with a non-zero precessional amplitude.
The marginalized  $\langle \theta_L \rangle$ posterior is constrained away from all three prior distributions, which can be accounted for by the relatively high network signal-to-noise ratio (SNR) of the event ($\ssim 19$) leading to better parameter constraints of additional quantities beyond $m_{1,2}$ and $\chieff$. 

Similarly to $\langle \theta_L \rangle$, the frequency $\langle \Omega_L \rangle$
also has a broad uninformative prior distribution. The mass conditioned prior prefers smaller $\langle \Omega_L \rangle$ values, confirming the near-linear relationship between low mass ratio values and $\langle \Omega_L \rangle$ explored in Ref.~\cite{2021PhRvD.103l4026G}. 
 Conditioning the prior on $\chieff$ skews it back to the middle and makes it nearly identical to the posterior distribution.

The behavior of the nutation frequency $\omega$  
is qualitatively similar to that of $\langle \Omega_L \rangle$: the posterior distribution is almost fully described by the information carried by the massed and effective spin.
Conditioning on the masses gives a distribution that prefers higher values of $\omega$ compared to the uninformative prior because, in general, lower values of $q$ correspond to higher values of $\omega$~\cite{2021PhRvD.103l4026G}.

For GW190412, the posterior and prior distributions of the nutational parameters $\Delta\theta_L$ and $\Delta\Omega_L$ are largely compatible,
a result we observe for most events across the entire dataset.
Nutations are a two-spin effect and as such they are intrinsically harder to measure~\cite{2014PhRvL.112y1101V,2016PhRvD..93h4042P,2021PhRvD.103f4067G}.

\subsection{Catalog constraints}

Our full results are reported in Table~\ref{eventtables} of Appendix~\ref{app}, %
where we list medians and $90$\% symmetric credible intervals of the uninformative prior, the conditioned priors, and the posterior for all our estimators ($\langle \theta_L \rangle, \langle\Omega_L\rangle, \Delta\theta_L, \omega$, $\Delta\Omega_L$, and the spin morphologies) for each BBH event in the current catalog. We see across our table that the posterior values typically have narrower credible intervals compared to their prior counterparts, compatible with the non-zero Hellinger distances between the uninformative priors and posteriors for each parameter. As is common practice in the field, we use equal-tailed credible intervals, which, for bound parameters,  exclude the extrema by definition. For events with high SNR, such as GW190412 and GW190814, the $90$\% credible intervals decrease significantly between prior and posterior (from widths of $\ssim 6$ radians to $\ssim 2$ radians) for well-measured parameters such as $\langle \theta_L \rangle$, tracing information gain from measurements.

\begin{figure*}[t] %
    \includegraphics[width = \textwidth]{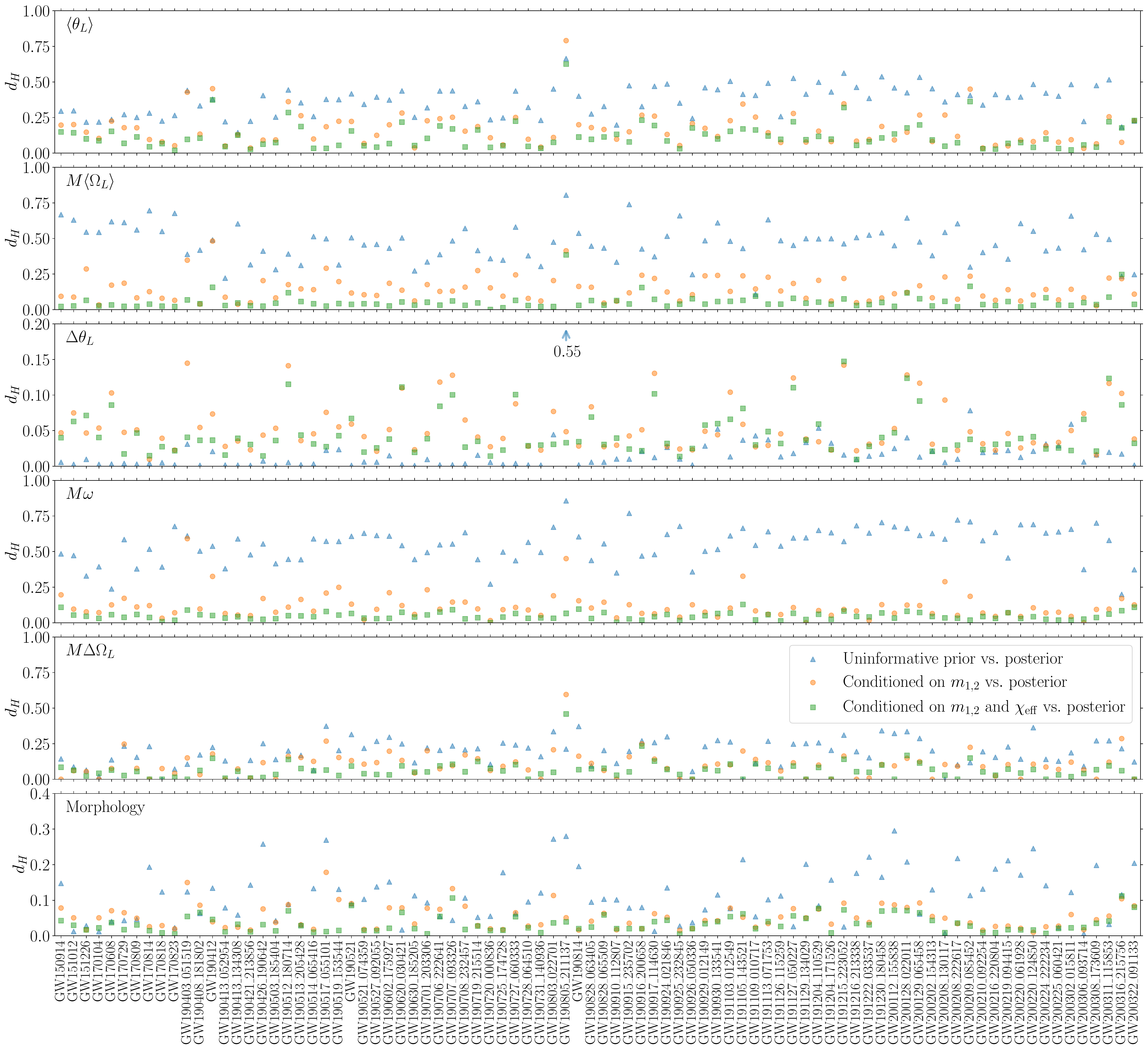}
    \caption{Hellinger distances $d_{H}$ for each of the five phenomenological parameters and spin morphologies comparing its three priors with the posterior for each GW event. 
    Scatter points indicate the distance between posterior vs. uninformative prior (blue triangles), posterior vs. prior conditioned on $m_{1,2}$ (orange circles), and posterior vs. prior conditioned on both $m_{1,2}$ and $\chieff$ (green squares).
   Note that the third from the top panel for $\Delta \theta$ and the bottom panel for the morphologies are scaled differently, reflecting the smaller $d_H$ values; GW190814 is an outlier with a distance between posterior and uninformative prior of $ \ssim 0.55$ (arrow).}
    
    \label{fig:fig4}
\end{figure*}

In Fig.~\ref{fig:fig4}
we summarize the Hellinger distances 
between the marginalized prior and posterior distributions of the various estimators. As expected, we find that the distance $d_H$  
decreases as the conditioning becomes stricter such that the posteriors and conditioned priors approach each other. Events where this is not the case, such as the Hellinger distances for $\Delta \Omega_L$ for the event GW200210\_092255, are those where the prior and posterior distributions are broad, leading to $d_H$ measurements that overestimate the differences between the distributions. The Hellinger distances for the nutational amplitude $\Delta \theta_L$ are small compared to the other four parameters, and conditioning the priors does not affect the $d_H$ values of the events (except for GW190814, whose well measured mass ratio significantly constrain the posterior of $\Delta \theta_L$). %

The only parameters where the distance $d_H$  between the posterior and the prior conditioned on $m_{1,2}$ and $\chieff$ is $>0.35$ for some events are $\langle \theta_L \rangle$ and
$\langle \Omega_L \rangle$. This should not be surprising: precession does not require spin-spin couplings and is thus easier to measure than nutation. The nutation frequency $\omega$ also shows cases with  $d_H>0.25$, but only between the uninformative prior and the posterior. Much like $\langle \theta_L \rangle$, the frequency $\omega$ has a strong dependence on the mass ratio $q$, but for this parameter the leading PN order does not depend on the BBH spins~\cite{2021PhRvD.103l4026G}. Therefore, the distance $d_H$ of the $\omega$ marginals decreases substantially once the priors are conditioned on the mass parameters. %

 The event GW190814 has the highest Hellinger distance $d_H$ values
 between its uninformative prior and posterior for almost all estimators. Its high network SNR of $25$ leads to tight constrains on the masses and spins of the two objects, which in turn meant that conditioning our prior on these tightly constrained quantities gave significant changes in the distributions and large distances.
 In particular, we report  
 $d_H = 0.44$ for $\langle \theta_L \rangle$,
$d_H = 0.65$ for $\langle \Omega_L \rangle$, 
 $d_H = 0.30$ for $\Delta \theta_L$,
 $d_H = 0.73$ for $\omega$,
$d_H = 0.05$ for $\Delta \Omega_L$,
  and 
$d_H = 0.28$ for the spin morphology,
  see Table~\ref{eventtables}. %
  GW190814 is the only event with a $d_H$ measurement above $0.2$ between the uninformative prior and posterior for the nutational amplitude $\Delta \theta_L$; however, its low mass ratio of $q = 0.11_{-0.01}^{+0.01}$ and spin posteriors constrain it to have negligible spin precession and nutation. %

\begin{figure*}[ht]
    \centering
    \includegraphics[width = 0.85\textwidth]{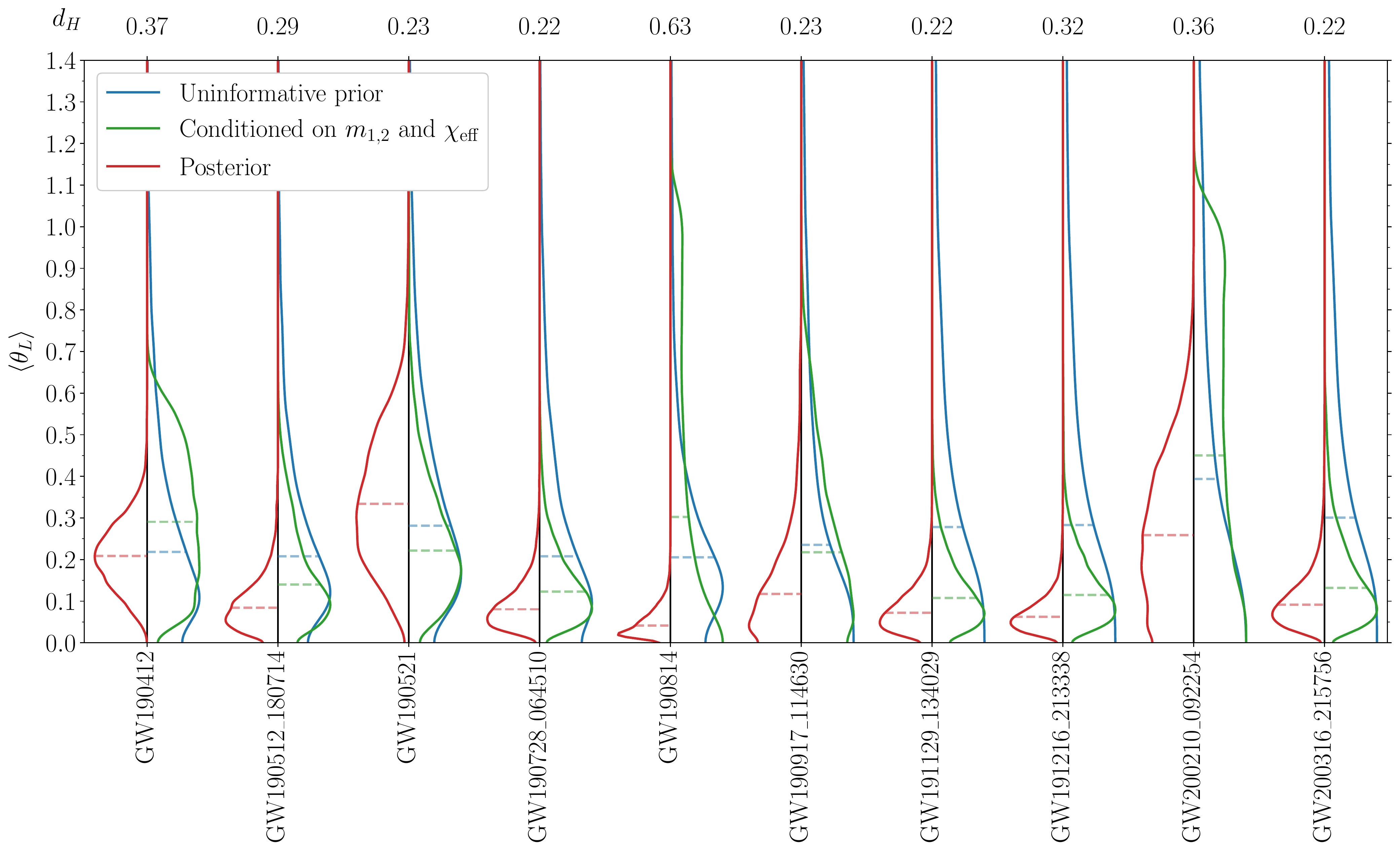}
    \caption{Precessional amplitude for the ten GW events with the largest Hellinger distance $d_H$ value between the posterior and prior conditioned on both masses and effective spins. Posteriors (red) are shown on the left-hand side of each violin plot, while the uninformative (blue) and conditioned (green) priors are shown on the right-hand side. Dashed lines indicate the median values of the corresponding distributions. The Hellinger distance $d_H$ value between the posterior and prior conditioned on both masses and effective spin is quoted above each event. }
    \label{fig:fig5}
\end{figure*}
Informed by the $d_H$ values listed in Table~\ref{eventtables} and Fig.~\ref{fig:fig4}, we select the ten events with the highest  $\langle \theta_L \rangle$ Hellinger distances between the posterior and the prior conditioned on both $m_{1,2}$ and $\chieff$. These are highlighted in  Fig.~\ref{fig:fig5}. Out of this subset of events, only GW190521 has a $\langle \theta_L \rangle$ posterior that prefers larger values compared to the conditioned prior. GW190521 is an event with high masses ($m_1=95.3^{+28.7}_{-18.9}$, $m_2=69.0^{+22.7}_{-23.1}$), contributing to a high network SNR ($\rho\sim 14.2$). The high network SNR
 leads to better parameter estimation and thus better-constrained posteriors. %
While its effective spin 
was measured to be compatible with $0$ ($\chieff = 0.03_{-0.39}^{+0.32}$), 
meaningful constraints on the spin misalignments %
led to claims of spin precession, quantified by an estimate of $\chip = 0.68_{-0.37}^{+0.25}$~\cite{2020PhRvL.125j1102A}. Evidence of spin precession for GW190521 persists when $\chip$ is generalized to include all variation over the precession timescale, $\langle \chip \rangle = 0.70_{-0.46}^{+0.56}$~\cite{2021PhRvD.103f4067G}. Similarly to GW190521, all of the events in Fig.~\ref{fig:fig5} but GW200210\_092254 and GW190917\_114630 are reported to have network SNRs $\gtrsim 10$. The lower SNR of GW200210\_092254 leads to wider prior and posterior distributions. Like for GW190814 and GW190412, the low mass ratio ($q=0.12^{+0.05}_{-0.05}$) of GW200210\_092254 leads to a conditioned prior that prefers a large precessional amplitude, while the posterior is somewhat constrained away from large $\langle \theta_L \rangle$; although the lower SNR means that we are unable to place an upper bound on the precession of this event unlike the ones placed for GW190814 and GW190412 (cf. Table~\ref{eventtables}). The event GW200129\_065458 has the largest median $\chip$ value in the GW catalogs \cite{2021arXiv211103606T}, but after conditioning our priors on just the masses, we do not find significant constraints placed on spin precession or nutation for this event.

\begin{figure*}
    \centering
    \includegraphics[width = 0.7\textwidth]{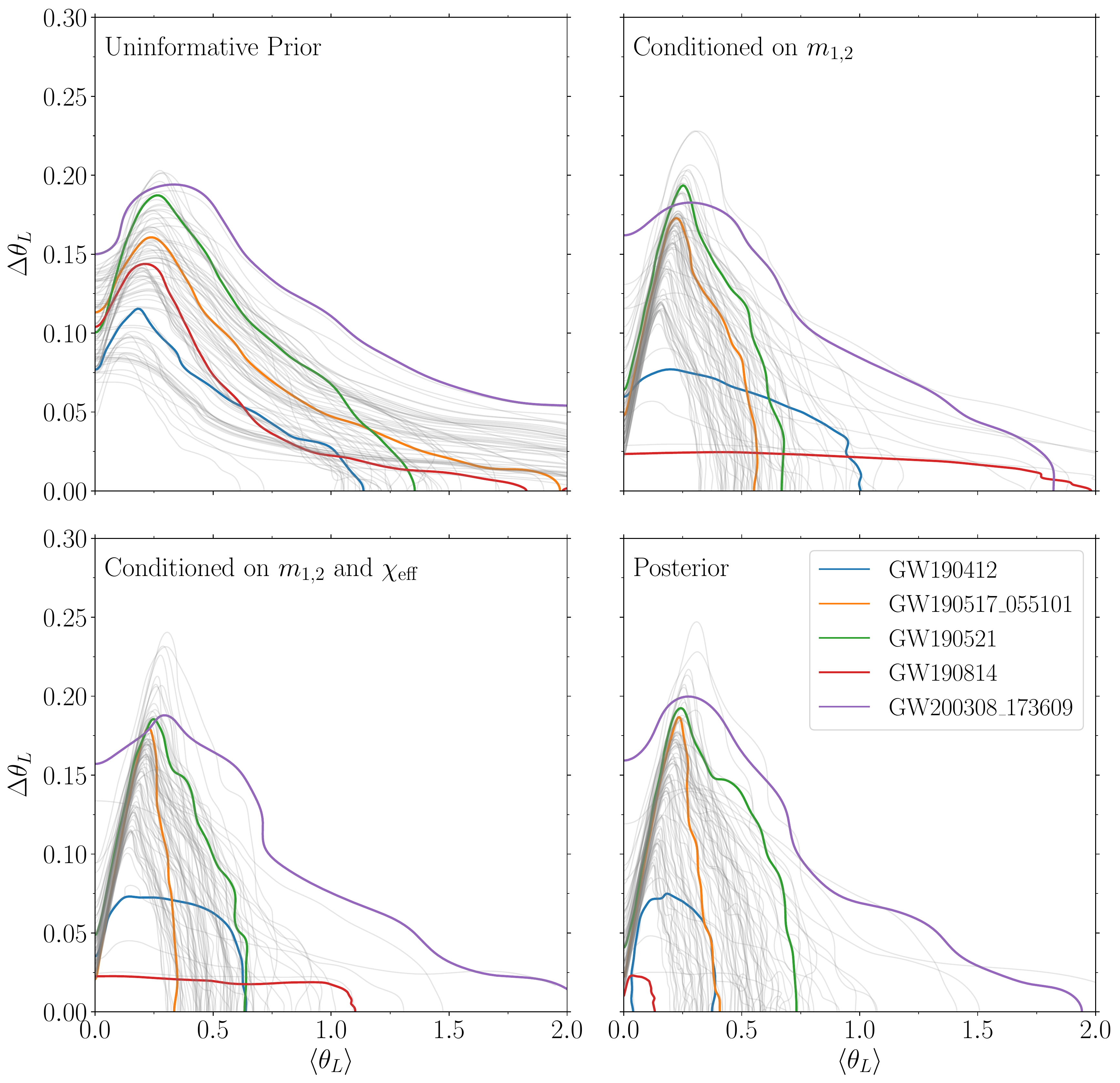}
    \caption{Joint prior and posterior marginalized distributions for the precession amplitude $\langle \theta_L \rangle$ and the nutational amplitude $\Delta \theta$ for all gravitational wave events, quoted at the $90$\% credible level. Some events with high Hellinger distance for either of the two amplitudes are highlighted in color.
    }
    \label{fig:fig6}
\end{figure*}

Figure \ref{fig:fig6} shows two-dimensional priors and posteriors for the precessional amplitude $\langle \theta_L \rangle$ and the nutational amplitude  $\langle \Delta\theta_L \rangle$ for all BBHs in the catalog. Colors highlight some events that may be of specific interest. Overall, we find that $\langle \theta_L \rangle$ is better constrained than $\Delta \theta_L$ for all of the events, highlighting once more that nutation is harder to measure than precession. In particular, all events are consistent with a nutational amplitude of $0$ at $90$\% credible interval. Unlike the nutational amplitude, most of the posteriors for $\langle \theta_L \rangle$ are constrained away from zero, something that is seen best with GW190412, cf. Fig.~\ref{fig:fig3} above. Current constraints on the nutational amplitude are overall poor, and tend to exclude high values ---this is best shown by the event posteriors of systems with large SNR and small $q$ such as GW190412 and GW190814. These constraints on the nutational amplitude are explained by the mass parameters of the events, as the difference between the posterior and the prior conditioned on the masses and the prior conditioned on $\chieff$ and the masses is negligible ($d_H < 0.15$). The posterior of GW190521 does not show the same behavior in the nutational amplitude because its less extreme mass ratio constrains it away from the single-spin limit that forbids nutations~\cite{2021PhRvD.103l4026G,2017CQGra..34f4004G}, and thus makes large nutational amplitudes possible.
The event GW200308\_173609 has a low SNR of $\sim 7.1$, and as a consequence its posterior distribution does not move away from its prior even after they are conditioned on the masses and the $\chieff$ of the event. In general, the gray lines representing the rest of the GW events have posterior distributions that constrain the precessional amplitude $\langle \theta_L \rangle$ to be smaller then the distributions given by the uninformative prior. For most of the population, the nutational amplitude posterior distribution remains unconstrained, in agreement with the rest of our findings.

\begin{figure*}[p] %
    \centering
    \includegraphics[width = \textwidth]{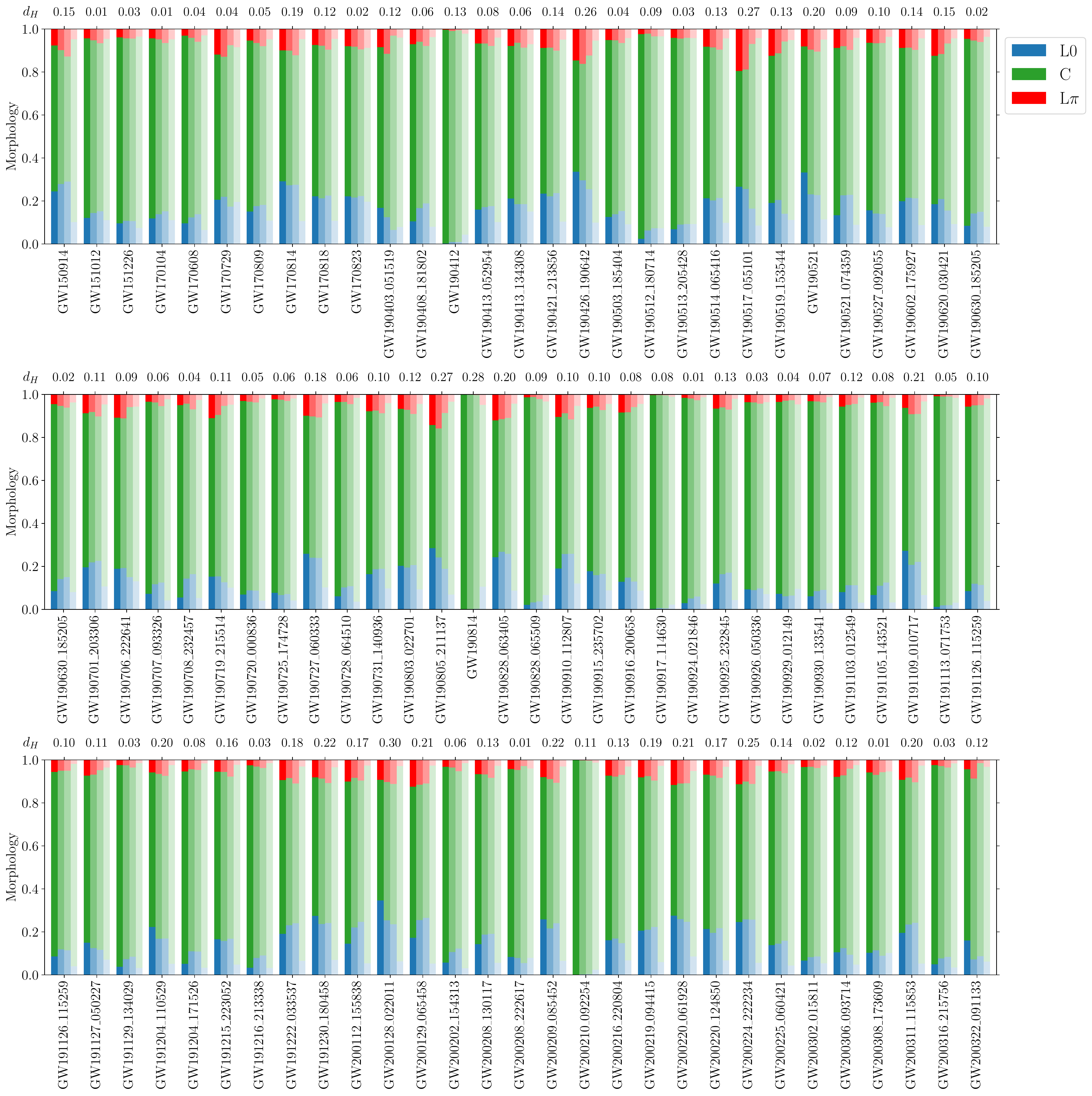}
    \caption{The fraction of samples in each of the three spin precession morphologies ---librating about $0$ (L$0$, blue), librating about $\pi$ (L$\pi$, red) and circulating (C, green)--- for all GW events. For each event, the fractions in the posterior distribution are shown by the most opaque, leftmost bar, followed by the fractions in each morphology for the prior conditioned on the masses and the effective spin $\chieff$, then the fractions for the prior conditioned on the masses distribution, and finally the fractions in the uninformative prior to the left. Above each event, we quote the Hellinger distance between the fractions in the uninformative prior distribution and the fractions in the posterior.  }

    \label{fig:fig7}
\end{figure*}

\begin{figure*}
    \centering
    \includegraphics[width = \textwidth]{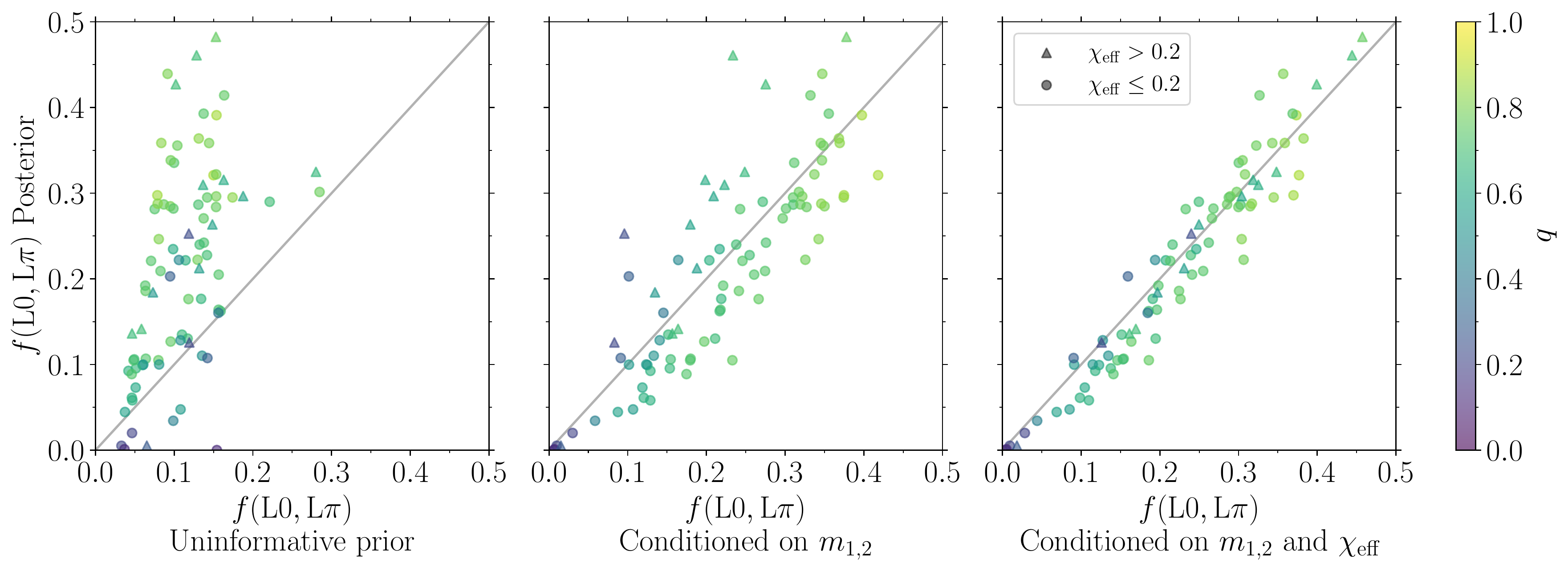}
    \caption{The fraction of samples  $f(\mathrm{L}0,\mathrm{L}\pi)$ in either of the two librating morphologies for all three priors against that for the posterior, progressively increasing the conditioning from left to right. Events are color-coded by the median value of their mass ratio $q$. The diagonal gray line indicates the case where the fraction of librating samples is the same for both distributions. Events with a median effective spin parameter $\chieff$ posterior value higher than $0.2$ are represented by triangles, while the rest are shown as circles.}
    
    \label{fig:fig8}
\end{figure*}

\begin{figure}
    \centering
    \includegraphics[width = \columnwidth]{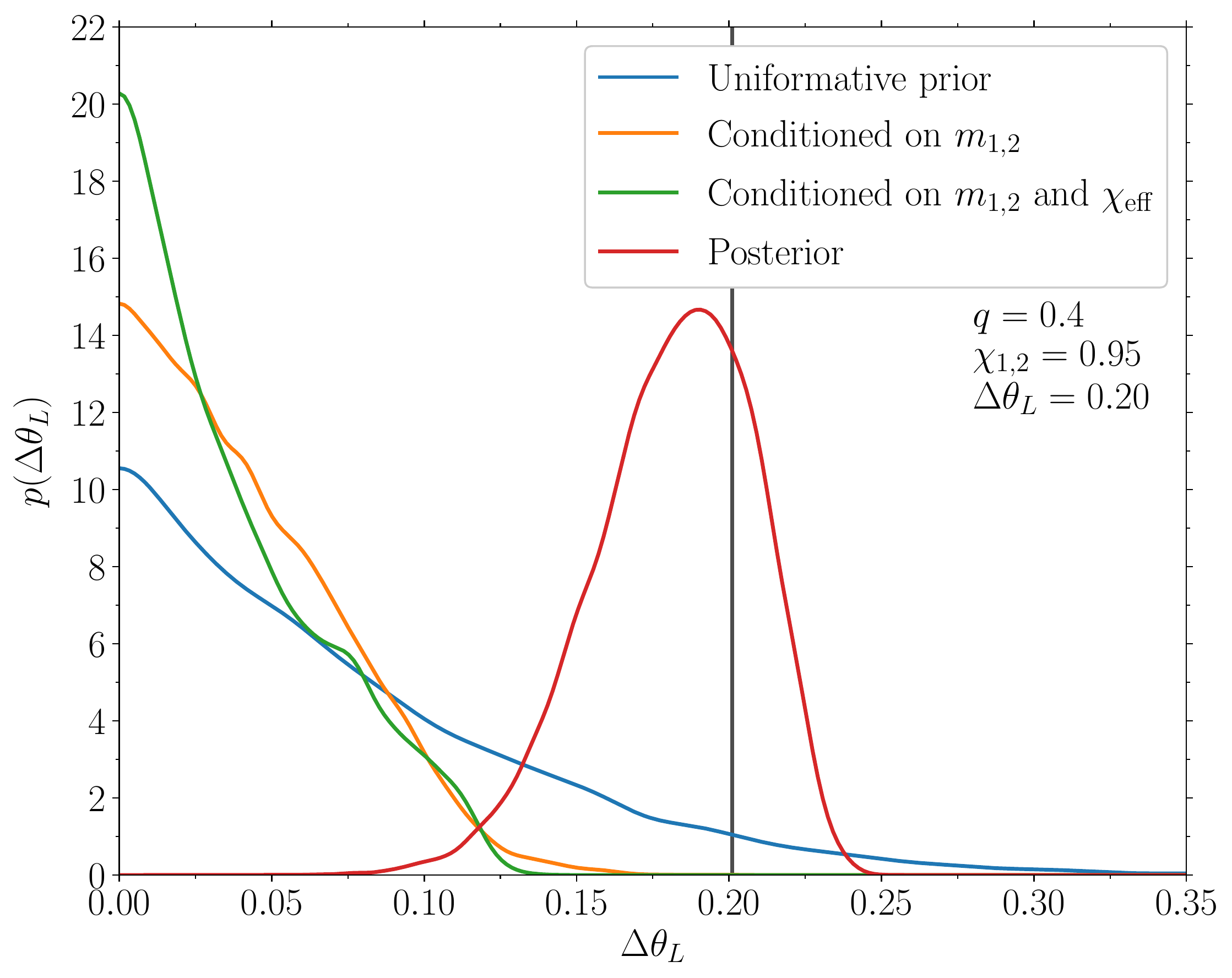}
    \caption{Distribution of nutational amplitude $\Delta \theta_L$ for a synthesized signal designed to maximize the nutational amplitude. The posterior (red) is well constrained from all the prior distributions (blue, orange, green), even when these are conditioned on masses and effective spin. The vertical black line represents the injected $\Delta \theta_L$ value.
    }
    \label{fig:fig9}
\end{figure}

\subsection{Spin morphologies}

We now look at constraining the BBH spin morphology~\cite{2015PhRvL.114h1103K,2015PhRvD..92f4016G}. Our results are presented in Table~\ref{eventtables} and Fig.~\ref{fig:fig7}.

 Events with mass ratios that significantly depart from unity (e.g. GW190412, GW190814) are constrained to be fully in the circular morphology. This is because the parameter space available to binaries in the two librating morphologies shrinks rapidly as $q\to0$~\cite{2015PhRvD..92f4016G}. Some events present  $d_H\gtrsim0.2$ between their uninformed priors and posteriors, but overall, we find that the morphology of a given event is largely determined once we condition on both masses and effective spin. For events with near equal masses, the uniform priors show an initial preference for the circulating morphology, which is then constrained to be smaller in their posteriors. %

Figure \ref{fig:fig8} illustrates the sensitivity of the spin morphologies to sequential prior conditioning. When comparing the fraction of posterior samples in any of the two librating morphologies (L0, L$\pi$) to the same fraction in the uninformed prior (left panel), for most events these fractions are constrained to be dissimilar, indicating that the data indicate that binaries are somewhat compatible with libration. As expected~\cite{2015PhRvD..92f4016G}, the fraction of librating samples in the posterior is closely correlated with the mass ratio (color scale).  Events with mass ratios below (above) approximately $0.5$ present a larger (smaller) fraction of the librating samples in the uninformed prior compared to the posterior. 
Once we condition the prior samples on the masses only (middle panel), for most events the probability of libration is approximately the same for the conditioned prior and the posterior. A similar correlation exists between the fraction of librating samples in the posterior and the events' $\chieff$. When the event priors are conditioned also on $\chieff$ (right panel), we see that events with $\chieff > 0.2$ in the posterior (depicted by the triangles) are pushed to the right, implying more event samples are librating in the prior conditioned on both the masses and $\chieff$ than the prior conditioned on the masses alone. Once both of these correlations are taken into account, the spin morphologies are fully described by their mass and $\chieff$ measurements, and no outliers remain in the right panel.

\section{Synthetic observation}
\label{injection}

Our investigations show that, overall, the SNR of the current GW catalog is too moderate to draw accurate constraints on our precession/nutation estimators. We now present a pilot study on synthetic data, showcasing the potential of a putative ``golden'' event for spin dynamics ---a hopeful prediction for the upcoming LIGO/Virgo/KAGRA observing run.

We fine-tune the parameters of a BBH such that spin nutations are manifestly prominent. In particular, we set $q = 0.4$ and spins of magnitude $\chi_{1,2} = 0.95$ directed into the orbital plane ($\theta_{1,2} = \pi/ 2$) at a reference frequency of $20$Hz. The angle between the two black hole spins in the orbital plane is set as $\Delta \phi = 0.1$. This results in an injected signal with $\langle \theta_L \rangle=0.59$, $M\langle\Omega_L\rangle=1.7\times10^{-3}$ (i.e. $4.6$ Hz in the detector frame), $\Delta\theta_L=0.20$, $M\omega=1.5\times 10^{-3}$ ($4.0$ Hz in the detector frame), and $M\Delta\Omega_L=4.8\times10^{-4}$  ($1.3$ Hz in the detector frame), and belongs to the circulating morphology.

The total source-frame mass of the system is set to $70 M_{\odot}$ to maximize the number of GW cycles in band, and
the orbital-plane inclination is $\simeq 30^\circ$, close to face-on. The source is placed at a %
luminosity distance of $500$ Mpc; the sky location is $({\rm RA, DEC}) = (0.75,0.5)$. We assume noise curves for LIGO and Virgo that are representative of the predicted detector performances during the upcoming the O4 observing run~\cite{2018LRR....21....3A}. Signals are injected and recovered using the IMRPhenomXPHM waveform model~\cite{2021PhRvD.103j4056P}. The injected signals have SNRs of about $45$ in LIGO Livingston, $33$ in LIGO Hanford, and $25$ in Virgo. We sample the resulting posterior using parallel nested sampling~\cite{2020MNRAS.498.4492S} as implemented in the {\sc Bilby} pipeline~\cite{2019ApJS..241...27A}, assuming  their standard uninformative priors.

Our full results are reported in Appendix~\ref{injapp}.  The distributions of the nutational amplitude $\Delta \theta_L$ is highlighted in Fig.~\ref{fig:fig9}.  In particular, the posterior is well-constrained from all three prior distributions, providing a confident detection of spin nutation. More specifically we find a
$d_H=0.55$ between the posterior and the uninformative prior,
a $d_H=0.89$ between the posterior and the prior conditioned on $m_{1,2}$,
and
a $d_H=0.95$ between the posterior and the prior conditioned on both $m_{1,2}$ and $\chieff$. Unlike for the GW events in the dataset, as we condition our priors on the masses and then on the masses and $\chieff$, the distributions are pushed towards low $\Delta \theta_L$ values. This is due to the priors on the spin magnitudes preferring low values, while the injected source has $\chi_{1,2} = 0.95$. Since we do not condition our priors directly on the spin magnitudes but only on $\chieff$, all our distributions but the posteriors present many samples with low $\chi_{1,2}$. On the other hand, nutations require high spins, thus pushing the posterior towards high $\Delta \theta_L$ values.

From our posterior distribution we report amplitudes $\langle \theta_L \rangle=0.59_{-0.04}^{+0.04}$, $M\langle\Omega_L\rangle=1.7_{-0.1}^{+0.1} \times 10^{-3}$, $\Delta\theta_L=0.19_{-0.05}^{+0.04}$, $M\omega=1.55_{-0.01}^{+0.01}\times 10^{-3}$, and $M\Delta\Omega_L=4.5_{-1.3}^{+1.1}\times 10^{-4}$, and a $0.0007$/$0.9993$/$0.0$ fraction of samples in the L0/C/L$\pi$ morphology, respectively. The prior and posterior distributions of all five precessional parameters are shown in Fig~\ref{fig:figapp}.

While limited to a single case, this exercise serves as a proof of concept, indicating that direct detection of spin nutations is not out of reach. For such a favorable event, our estimators are indeed sensitive to additional information beyond the masses and the dominant spin parameter~$\chieff$.

\section{Conclusions}
\label{sec:conclusions}

Spin precession and nutation are both subtle effects on the waveform. While statements such as \emph{``at least one of the component black holes has spin greater than 0.2''} were possible since the very first GW detections~\cite{2016PhRvL.116x1103A}, detailed spin dynamics is much harder to extract from the data. In this paper, we exploited previously developed estimators of BBH spin precession and nutation~\cite{2021PhRvD.103l4026G, 2015PhRvL.114h1103K, 2015PhRvD..92f4016G} in conjunction with current GW data.  

Precisely because spin effects are subdominant, we tackled the interpretation issue on whether indirect constraints from other, easier-to-see parameters are enough to explain features in the data. To this end, we formalized and systematically applied a sequential prior-conditioning approach.

Our analysis does not find strong evidence of either precession or nutation in any individual event using our phenomenological estimators. These results are compatible with those of Refs.~\cite{2019PhRvX...9c1040A, 2021arXiv210801045T, 2021PhRvX..11b1053A, 2021arXiv211103606T}, which also find no compelling evidence for spin precession in single events. Sequential prior conditioning indicates that, while comparisons between uninformative priors and posteriors could be used to claim evidence for spin precession, their differences are largely re-absorbed when one takes into account measurements of masses and effective spins.

We also presented a pilot injection study, proving  that favorable sources at current sensitivities will indeed allow us to disentangle precession and nutation from the coupled motion of the BBH spins. This showcases the potential of our spin estimators parameters to uncover finer details from GW signals, paving the way to deeper explorations in terms of both fundamental physics and astronomy~\cite{2018PhRvD..98h4036G,2019PhRvD..99j3004G,2021MNRAS.501.2531S,2022arXiv220600391S}.

\vspace{-0.5cm}

\acknowledgements

We thank Matthew Mould and Cameron Mills for discussions.
D.\,Gangardt, D.\,Gerosa, V.D.R., and N.S. are supported by European Union's H2020  ERC Starting Grant No.~945155--GWmining, Cariplo Foundation Grant No.~2021-0555, and Leverhulme Trust Grant No.~RPG-2019-350.
D.\,Gangardt acknowledges partial support by the `Study in Italy' program of the Italian Ministry of Foreign Affairs and International Cooperation. 
M.K. is supported by the National Science Foundation Grant No. PHY-1607031 and No. PHY-2011977.
V.D.R. acknowledges partial support by the ERC H2020 project HPC-EUROPA3 (INFRAIA-2016-1-730897). 
Computational work was performed on the University of Birmingham BlueBEAR cluster and at CINECA with allocations through INFN, Bicocca, and ISCRA Type-B project HP10BEQ9JB. %

\appendix

\vspace{-0.3cm}

\section{Precessional and nutational parameters of the synthetic signal}
\label{injapp}
\begin{figure*}[t]
    \centering
    \includegraphics[width = 0.8\textwidth]{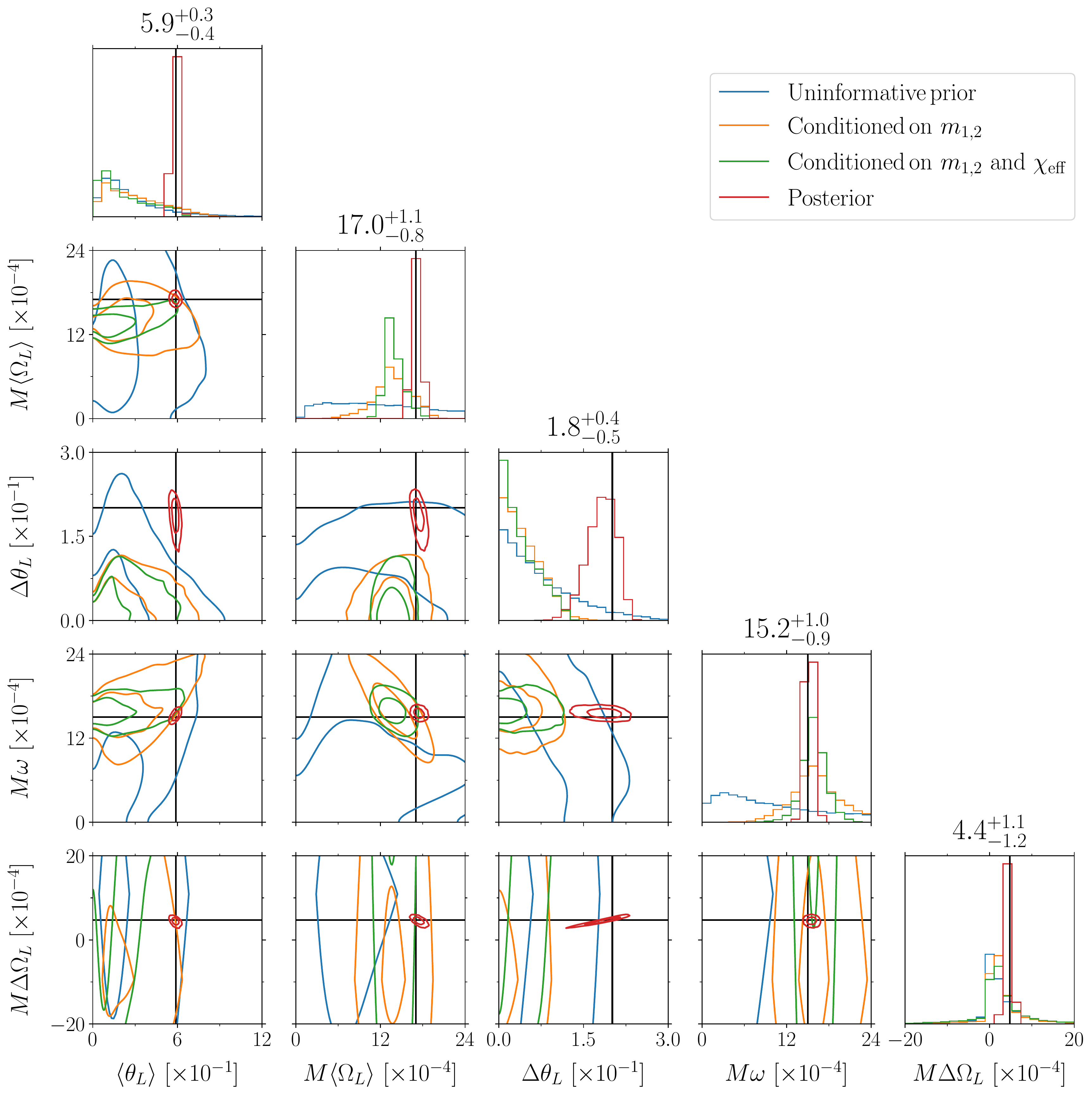}
    \caption{Uninformative prior (blue), prior conditioned on the masses only (orange), prior conditioned on the masses and on $\chieff$ (green) and posterior (red) distributions of the five precessional and nutational parameters for the synthetic event described in Sec.~\ref{injection}. The joint distributions represent the 90\% and 50\% confidence levels. The black vertical and horizontal lines show the injected system. The median value of the posterior for each parameter is displayed above the marginalized distributions.
    }
    \label{fig:figapp}
\end{figure*}

\vspace{-0.3cm}
Figure~\ref{fig:figapp} presents the distributions of the five precessional and nutational parameters of the synthetic observation described in Sec.~\ref{injection}. The injected value of each parameter is successfully recovered within the 90\% confidence intervals of the posterior distribution.

\vspace{-0.3cm}

\section{Full results in tabular form}
\label{app}
\vspace{-0.3cm}

Table~\ref{eventtables} reports results for all distributions and all our estimators across the current GW catalog. %

\vspace{-1cm}

\clearpage

\input{table_flipped}

$\,$\\$\,$\\$\,$\\$\,$\\$\,$\\$\,$\\$\,$\\$\,$\\$\,$\\$\,$\\$\,$\\$\,$\\$\,$\\$\,$\\$\,$\\$\,$\\$\,$\\$\,$\\$\,$\\$\,$\\$\,$\\$\,$\\$\,$%

\bibliography{bibme}

\end{document}

%% file: table_flipped.tex
\setlength{\LTcapwidth}{\textwidth}
\renewcommand{\arraystretch}{0.8}
\setlength{\tabcolsep}{2.5pt}
\begin{longtable*}{
@{$\!\!\!\!\!\!\!\!\!\!\!\!\!\!$}c
>{\footnotesize}c
|
>{\footnotesize}c
||
>{\footnotesize}c  
|
>{\footnotesize}c  
|
>{\footnotesize}c 
|
>{\footnotesize}c
|
>{\footnotesize}c 
|
>{\footnotesize}c
}
\label{eventtables}
& \bf Event & \bf Prob. & \bf $\langle \theta_L \rangle [\times 10^{-1}]$ & \bf $M\langle \Omega_L \rangle [\times 10^{-4}]$ & \bf $\Delta \theta_L  [\times 10^{-1}]$ & \bf $M\omega  [\times 10^{-4}]$ &
\bf $M\Delta \Omega_L  [\times 10^{-4}]$ &
L$0$/C/L$\pi$ \\[3pt]
\hline
 & & & & & & & & \\[-6pt]
\hline
 & & & & & & & & \\[-8pt]
& GW150914 & Uninformative Prior & $1.8_{-1.4}^{+5.7}$&  ${13.8}_{-{10.1}}^{+{20.4}}$& ${2.7}_{-{2.5}}^{+{6.7}}$& $0.8_{-0.7}^{+{4.4}}$& $6.7_{-{69.0}}^{+{125.8}}$& ${0.1}/{0.85}/0.05$ \\&& Cond. on $m_{1,2}$ & $1.6_{-1.2}^{+2.0}$ &  ${17.9}_{-{4.0}}^{+3.9}$ & $2.4_{-2.3}^{+{7.8}}$ & $0.4_{-0.2}^{+0.4}$ & $0.3_{-{43.5}}^{+{51.1}}$ & $0.29/0.58/0.13$ \\&& Cond. on $m_{1,2}$ and $\chieff$ & $1.5_{-1.2}^{+{1.9}}$ &  ${17.8}_{-{3.4}}^{+{3.5}}$ & $2.3_{-2.2}^{+{7.9}}$ & $0.4_{-0.3}^{+{0.3}}$ & $0.5_{-{39.7}}^{+{51.0}}$ & $0.28/0.62/0.1$ \\&& Posterior& $1.1_{-0.9}^{+1.8}$ &  ${17.7}_{-{3.2}}^{+{3.4}}$ & $2.0_{-1.9}^{+7.3}$ & $0.3_{-0.2}^{+0.4}$ & $0.6_{-{47.1}}^{+{54.8}}$ & $0.24/0.68/0.08$ \\
\hline
 & & & & & & & & \\
\end{longtable*}